\documentclass[12pt,a4paper]{article}
\usepackage[final]{showkeys}
\usepackage{amsmath}
\usepackage{amssymb}
\usepackage{amsthm}
\newcommand{\Eins}{\boldsymbol 1}

\renewcommand{\t}[1]{\texttt{#1}}

\newcommand{\tr}{\mathop\mathrm{Tr}\nolimits}

\title{Quantum error correction fails for Hamiltonian models}
\author{R.~Alicki~\footnote{Electronic address: 
\t{fizra@univ.gda.pl}} \\
{\normalsize Institute of Theoretical Physics and Astrophysics} \\
{\normalsize University of Gda\'nsk, Poland}} 
\begin{document}

\maketitle
\begin{abstract}
It is argued that the existing schemes of fault-tolerant quantum computation
designed for discrete-time models and based on quantum error correction fail for continuous-time Hamiltonian
models even with Markovian noise.
\end{abstract}
The fundamental challenge to quantum information processing is a devastating influence of decoherence
processes due to the interaction of a quantum device with environment. It is rather generally believed that,
at least in principle, by suitable active procedures of {\em quantum error correction} this problem can be solved if the level
of noise is lower then a certain threshold. The aim of this note is to show that, unfortunately, the success of existing error correction procedures is due to the discrete in time modelling of quantum evolution.
In physical terms discrete models correspond to unphysical infinitely fast gates. More precisely we shall argue that the fidelity of the final state of the quantum computer is of the order of 
\begin{equation}
{\rm fidelity} \simeq \exp(- \lambda^2 t_0 V)
\label{fid}
\end{equation}
where $\lambda^2$ is the decoderence rate for a single qubit, $t_0$ is the duration of a unitary gate and $V$ is the
"volume" of an algorithm.
As we are going to construct a counterexample it is enough to discuss the idea of {\em fault-tolerant quantum information 
processing} for the simplest model of protecting a bit of classical information for a given, but arbitrarily long period of time and with an arbitrarily high accuracy. The extension of this results to a general fault-tolerant quantum computation is straigthforward. The presented framework is general enough to cover all known standard examples of fault-tolerant procedures.\\
The mathematical model consists of a $M$-qubit quantum register and a large enough set $A$ of ancillary qubits prepared
in a standard state $|\phi_A> = |000...00>$. The state $|0>$ of an ancillary qubit is assumed to be stable with respect to noise. The controlled part of the evolution is governed by a sequence of unitary operations (1 or 2-qubit {\em quantum gates}) acting on the register and ancillas and executed in discrete time steps. Noise is modelled by completely positive trace preserving (error) maps  acting on the register and those ancillary qubits which are engaged in the process of error correction at a given time step. Any gate including trivial one is followed by a suitable error map. By a {\em code} we mean in our specific situation a 2-dimensional subspace of the register states spanned by the logical $|0_L>$ and $|1_L>$. We shall use $\|\cdot\|_1 ,\|\cdot\|_{\infty}$ and $ \|\cdot\|$ for tracial , operator and superoperator norms, respectively.\\
The dynamics corresponding to $T$ time steps without error corrections is simply given by a $(E)^T$ where $E$ is an error map acting on the register.
The error map $E$ is composed of elementary error maps $E^{(\alpha)}$ typically corresponding to 1 or 2 - qubit locations
and satisfying the estimation
\begin{equation}
\|E^{(\alpha)} - {\mathbb I}\| \leq p
\label{erp}
\end{equation}
where $p$ provides an overall bound on the error rate.

In the next step of construction we add after any error map at the time step $k$ a sequence of gates producing the unitary map $R_k$ which acts on the register and a subset of ancillas $A_k$ (different for any $k$) and which are designed to recover the input state. Then the evolution is modelled by a sequence of dynamical maps 
\begin{equation}
R_T E R_{T-1}E \cdots R_1 E \ .
\label{evo1}
\end{equation}
Obviously, the real gates are also accompanied by noise,  and should produce rather completely positive dynamical maps $R'_k$ than the unitaries $R_k$. We can always write $ R'_k = E'_k R$ and define $E_k = E E'_k$ with 
$E_0 \equiv E$. The complete evolution in our model is now a composition of maps
\begin{equation}
E'_TR_T E_{T-1} R_{T-1}E_{T-2} \cdots R_1 E_0 \ .
\label{evo2}
\end{equation}
The error maps $E_k$ and the unitary recovery maps $R_k$ should satisfy the following {\em error correction property} with the accuracy $\mu$
\begin{equation}
R_k E_{k-1}P_{\psi}\otimes P_{\phi_{A_k}}\otimes\rho_{k-1}= (1-\mu)P_{\psi}\otimes \rho_k + \sigma_k\ ,\ \|\sigma_k\|_1 \leq B\mu\ ,
\label{corr}
\end{equation}
with a certain fixed $B \geq 1$.
Here $\rho_k$ is an unspecified state of ancillas from $\bigcup_{l=1}^k A_l$, $P_{\psi}= |\psi><\psi|$ is any
pure state from the code and   $P_{\phi_{A_k}}=  |\phi_{A_k}><\phi_{A_k}|$. \\
One should notice that the condition (\ref{corr}) is more restrictive for $k>1$ than for $k=1$, as in the former
case more faults should be corrected and the propagation of them becomes very important.

We are interested in the "almost final" state 
of the system (prior to the last error map $E'_T$)
\begin{equation}
\rho_T^{af} =   R_T E_{T-1} R_{T-1}E_{T-2} \cdots R_1 E_0 P_{\psi}\otimes P_{\phi_A}\ .
\label{finst}
\end{equation}
Combining (\ref{finst}) with (\ref{corr})  we obtain 
a lower bound for the fidelity of the "almost final" state of the register
\begin{equation}
<\psi ,{\rm Tr}_A \bigl(\rho_T^{af}\bigr)\psi>\geq  (1-\mu)^T - B\bigl[1- (1-\mu)^T\bigr]\ .
\label{ine1}
\end{equation}
The protection of a single bit is succesfull if for any $T$ and any $\epsilon >0$ there exists encoding
into large enough $M$-qubit space together with recovery maps $R_k$ satisfying error correction property (\ref{corr})
with $\mu = \epsilon/(B+1)T$. Under these conditions, the fidelity of the "almost final" register's state is bounded from below by $(1-\epsilon)$ and one shows that using enough redundances and majority voting one can extract the classical bit even from this state perturbed by the last error map $E'_k$.\\
The remarkable and highly involved results \cite{ABO}\cite{KLZ}\cite{P}\cite{NC} show by explicite construction of concatenated codes and encoded recovery maps and for the error maps satisfying reasonable conditions that the presented above scheme works for the model of general quantum computation. The conditions concern locality, statistical independence and the {\em threshold condition} ($p< p_c$) imposed on elementary
error maps $E^{(\alpha)}$. Moreover, these procedures are {\em efficient} i.e. the number of register qubits and ancillas moderately grows like ${\rm poly}(\ln (1/\epsilon)$.\\

We shall try to apply the scheme of above to a simple model of continuous-time irreversible dynamics governed by
a Markovian master equation.
The strictly Markovian, single-qubit noise acting on the register is explicitely given in terms of the single-qubit semigroup generators
\begin{equation}
L_k\rho = \lambda^2 \bigl(\Phi_k \rho - \rho)\ ,{\rm where}\ \Phi_k = \frac{1}{2}{\tr}_k\rho\otimes{\Eins}_k \ ,\ k=1,2,...M\ .
\label{gen} 
\end{equation} 
In this model $\lambda^2$ is a single qubit decoherence rate and $\lambda^2 t_{clock}$ corresponds to $p$ in the discrete model where $t_{clock}$ is a time step of the computer's clock.\\ 
The total Hamiltonian of the register and ancilla  system is denoted by $H(t); t\in[0,T\tau]$ and is designed to execute the
sequence of unitary state recovery operations $R_TR_{T-1}\cdots R_1$. Here $\tau > t_{clock}$ is a working period needed to execute the continuous-time analog of the operation $R_kE_{k-1}$. The time-dependent Hamiltonian possesses the following structure
\begin{equation}
H(t) = \sum_{\alpha} f_{\alpha}(t)h^{\alpha} 
\label{gates}
\end{equation}
i.e. consists of well-separated but possibly parallel pulses $f_{\alpha}(t)h^{\alpha} , f_{\alpha}(t)\geq 0 , \int_{-\infty}^{\infty} f_{\alpha}(t) dt =1$, of the typical width $t_0<t_{clock}$. Here $h^{\alpha}$ are  1 or 2-qubit Hamiltonians acting on register and ancillas. For the rest of the discussion the model of noise acting on ancillas is irrelevant. Taking the limit $f_{\alpha}(t)\to \delta (t-t_{\alpha})$ and using a Markovian model of noise we obtain the standard discrete time model (\ref{evo2}).

We shall discuss the less demanding error correction property (\ref{corr}) for $k=1$ which can be written
in terms of fidelity
\begin{equation}
F_1 =<\psi ,{\rm Tr}_A \bigl(R_1E P_{\psi}\otimes P_{\phi_A}\bigr)\psi>\geq  \bigl(1-(B+1)\mu\bigr)\ .
\label{con1}
\end{equation}
This condition involves the error map acting on the register in the first time step combined with the {\em errorfree evolution} of the register and ancillas designed to recover the initial state $\psi$. Therefore the continuous time counterpart to the map $R_1E$ 
is the propagator $\Lambda $ obtained as a solution of the master equation in the first working period $[0,\tau]$ ($\hbar\equiv 1$)
\begin{equation}
\frac{d}{dt}\rho  = -i[H(t),\rho] +\lambda^2\sum_{k=1}^M \bigl(\Phi_k \rho - \rho)\ . 
\label{MME} 
\end{equation}   
Introducing the Hamiltonian (super)propagator for $t\geq s$
\begin{equation}
{\cal U}(t,s) = {\mathbb T}\exp\Bigl\{-i\int_s^t[H(u),\cdot ]du\Bigr\}\ ,\ {\cal U}(s,t)\equiv{\cal U}(t,s)^{-1}
\label{pro} 
\end{equation}   
we can write the propagator solving (\ref{MME}) in terms of the series expansion
\begin{equation}
\Lambda(t) = {\cal U}(t,0) \Bigl(e^{-\lambda^2Mt  }\sum_{N=0}^{\infty}(\lambda^{2N}\sum_{k_N,..,k_2,k_1=1}^M \int_0^t dt_N\int_0^{t_N} dt_{N-1}\dots\int_0^{t_2}dt_1 
\nonumber
\end{equation}
\begin{equation}
\Phi_{k_N}(t_N)\Phi_{k_{N-1}}(t_{N-1})\dots\Phi_{k_1}(t_1)\Bigr)
\label{sol} 
\end{equation}   
where
\begin{equation}
\Phi_k (u) = {\cal U}(u,0)^{-1}\Phi_k{\cal U}(u,0)\ . 
\label{map} 
\end{equation}
To estimate the fidelity of the protected state after the first working period $[0,\tau]$ it is sufficient to use (\ref{sol}) together with the {\em recovery condition}
\begin{equation}
{\cal U}(\tau,0)= {\mathbb I}\otimes {\cal U}_A(\tau,0)\ .
\label{rec} 
\end{equation}
Denote by $F(t), t\in[0,\tau]$ the following expression for the "time-dependent fidelity"
\begin{equation}
F(t) = \Bigl(e^{-\lambda^2 M t}\sum_{N=0}^{\infty}\lambda^{2N}\sum_{k_N,..,k_2,k_1=1}^M \int_0^t dt_N\int_0^{t_N} dt_{N-1}\dots\int_0^{t_2}dt_1 
\nonumber
\end{equation}
\begin{equation}
{\rm Tr}\bigl(P_{\psi}\otimes{\Eins}_A\Phi_{k_N}(t_N)\Phi_{k_{N-1}}(t_{N-1})\dots\Phi_{k_1}(t_1)P_{\psi}\otimes P_{\phi_A}\bigr)\Bigr)\ .
\label{fidt} 
\end{equation}     

From (\ref{fidt}) it follows that the "time-dependent error" $E(t)= 1-F(t)$ satisfies the equation
\begin{equation}
\frac{dE}{dt}= -\lambda^2m E(t) + \lambda^2m X(t)
\label{err} 
\end{equation}  
where
\begin{equation}
X(t) = \Bigl(e^{-\lambda^2 M t}\sum_{N=0}^{\infty}\lambda^{2N}\sum_{k_N,..,k_2,k_1=1}^M \int_0^t dt_N\int_0^{t_N} dt_{N-1}\dots\int_0^{t_2}dt_1 
\nonumber
\end{equation}
\begin{equation}
\Bigl[1-(\frac{1}{M}\sum_{k=1}^M{\rm Tr}\bigl([\Phi_k(t)P_{\psi}\otimes{\Eins}_A]\Phi_{k_N}(t_N)
\Phi_{k_{N-1}}(t_{N-1})\dots\Phi_{k_1}(t_1)P_{\psi}\otimes P_{\phi_A}\bigr)\Bigr]\Bigr)\geq 0\ .
\label{force} 
\end{equation}   
Using properties of norms we obtain
\begin{equation}
\frac{1}{M}\sum_{k=1}^M{\rm Tr}\Bigl(\bigl[\Phi_k(t)P_{\psi}\otimes{\Eins}_B\bigr]\Phi_{k_N}(t_N)
\Phi_{k_{N-1}}(t_{N-1})\dots\Phi_{k_1}(t_1)P_{\psi}\otimes P_{\phi_A}\Bigr)
\nonumber
\end{equation}
\begin{equation}
\leq
\max_k \|\Phi_k(t)P_{\psi}\otimes{\Eins}_A\|_{\infty}
\label{force1} 
\end{equation}   
and hence
\begin{equation}
X(t)\geq x(t) =
1-\max_k \|\Phi_k(t)P_{\psi}\otimes{\Eins}_A\|_{\infty}\geq 0\ .
\label{force2} 
\end{equation}  
Solving equation (\ref{err}) and using (\ref{force2}) we obtain the lower bound for the error at the end of the
working period
\begin{equation}
E(\tau) = \lambda^2 M\int_0^{\tau} e^{-\lambda^2M({\tau}-s)}X(s)\,ds \geq \lambda^2 M\int_0^{\tau} e^{-\lambda^2M({\tau}-s)}x(s)\,ds\ .
\label{bound} 
\end{equation}  
Now we have to analyze the time-dependence of the function $x(t); t\in[0,\tau]$. We notice first that due to
the condition (\ref{rec}) we have 
\begin{equation}
\Phi_k(0)=\Phi_k(\tau)=\Phi_k
\label{boX} 
\end{equation}  
what gives
\begin{equation}
x(0)=x(\tau) =1-\|\Phi_k P_{\psi}\|_{\infty}\geq \frac{1}{2}
\label{boX1} 
\end{equation}  
and the inequality follows from the simple estimation $\|\Phi_k P_{\psi}\|_{\infty}\leq 1/2$.
The function $x(t)$ is non-negative, continuous and larger then $1/2$ at the ends of the time interval $[0,\tau]$.

In principle one cannot exclude that using suitable controlling Hamiltonian
$H(t)$ we can make $X(t)$  arbitrarily  small in the interval $[\delta , \tau -\delta]$ with arbitrarily  small $\delta >0$. This could be achieved only by applying fast enough gates in the procedure of "hiding and recovering" of the initial state $P_{\psi}$. However, any technology applied to implement quantum information processing puts the limit on the speed of gates which can be given in terms of the typical width of a pulse 
\begin{equation}
t_0\geq t_0^{min}\ . 
\label{gs}
\end{equation}
This condition can be also expressed by the following inequality involving (super)operator norm
\begin{equation}
f_{\alpha}(t)\|[h^{\alpha},\cdot]\| \leq C(t_0^{min})^{-1}
\label{speed} 
\end{equation}  
where $C= {\cal O}(1)$ is a certain constant.\\
The completely positive map $\Phi_k$ can be written in the Kraus form
\begin{equation}
\Phi_k\rho = \frac{1}{2}\sum_{\mu,\nu= 0,1} e^{(k)}_{\mu\nu}\rho\, e^{(k)}_{\nu\mu}
\label{kraus} 
\end{equation}  
where $e^{(k)}_{\mu\nu}= |\mu><\nu|$ for $k$- qubit. 

Therefore, using (\ref{speed}),(\ref{kraus}) and the fact that only 1 or 2 qubit gates can be applied to a given register qubit in the time intervals $[0,t_0^{min}]$ and $[\tau-t_0^{min},\tau]$,  we can estimate
the time derivative of $x(t)$ 
\begin{equation}
|\frac{d}{dt} x(t)|\leq \max_k \|\frac{d}{dt}\Phi_k(t)P_{\psi}\otimes{\Eins}_B\|_{\infty}
\leq 4C t_0^{-1}\ ,\ {\rm for}\ t\in [0,t_0^{min}]\cup[\tau-t_0^{min},\tau]\ .
\label{sp1} 
\end{equation}
It implies that for $t\in [0,t_0^{min}]\cup[\tau-t_0^{min},\tau]$, $x(t)= {\cal O}(1)$ and determines the lower bound for the integral in (\ref{bound}). This leads to the
basic estimations of the fidelity at the end of the first working period in terms of the parameter $q\simeq \lambda^2 t_0^{min}$  
\begin{equation}
F(\tau)\equiv F_1^{cont} \leq 1 -  M q\ ,\  {\rm for}\  M q<<1 
\label{fid4}
\end{equation}
with the saturation for large $M$
\begin{equation}
F_1^{cont} \leq \frac{1}{2}\ ,\ {\rm for}\  M q>>1 \ .
\label{fid5}
\end{equation}
It follows that the inequality (\ref{con1}) with $\mu = \epsilon/(B+1)T$ cannot be satisfied and the same is true for all working periods. Therefore, a successful state protection is not feasible for real quantum systems, at least within
the existing schemes of error correction.

{\bf Conclusions}\\
Using a tractable continuous time model of quantum information processing with the {\em limited gate's speed condition}
(\ref{gs})  we conclude that during any working period the input state's fidelity is reduced by a factor given by (\ref{fid4})(\ref{fid5}). This loss of information cannot be reversed using multiqubit encodings and leads to an exponential decrease of the total fidelity with the volume of the algorithm (\ref{fid}). The crucial parameter is the duration of the gate $t_0$ multiplied by the decoherence rate $\lambda^2$. Only in the limit $t_0\to 0$ one can recover the results of the theory of fault-tolerant quantum computation. We obtain also a rough estimation
\begin{equation}
 T_{max} \simeq \frac{1}{\lambda^2 t_0^{min}}
\label{vol}
\end{equation}
for the maximal number of algorithm's steps which can be executed within a given physical implementation. This formula follows from (\ref{fid}) and the fact, that we can tolerate finite error per qubit at the end of quantum computation.\\
The presented results explain also the absence of the analogical 
uncorrectable errors in the classical theory of digital computers. Namely, for the classical digital case no continuous in time reversible operations are possible and therefore  $t_0=0$, intrinsically. However, continuous description of classical computers which takes into account the physically existing continuum of states between logical $\{0\}$ and $\{1\}$ will lead to the results similar to those in quantum theory. This fact can be important for the mesoscopic and microscopic realizations of classical computers in the future.

\textbf{Acknowledgements:} 
The author is extremely grateful to Micha\l\ , Pawe\l\ and Ryszard Horodecki for their countless stimulating and vigorous discussions concerning fault tolerant computation. The exchange of opinions with Daniel Lidar is acknowledged
also. This work was supported by the Polish Ministry of Scientific Research under Grant No PBZ-Min-008/P03/03. 

\end{document}